\documentclass[journal=ancac3,manuscript=article]{achemso}

\usepackage[version=3]{mhchem} 

\author{Sonia Melle}
\affiliation{Department of Optics, Complutense University of Madrid, E-28037 Madrid, Spain}
\email{smelle@fis.ucm.es}

\author{Oscar G. Calder\'{o}n}
\affiliation{Department of Optics, Complutense University of Madrid, E-28037 Madrid, Spain}

\author{Marco Laurenti}
\affiliation{Department of Chemistry in Pharmaceutical Sciences, Complutense University of Madrid, E-28040 Madrid, Spain}

\author{Diego Mendez-Gonzalez}
\affiliation{Department of Chemistry in Pharmaceutical Sciences, Complutense University of Madrid, E-28040 Madrid, Spain}

\author{Ana Egatz-G\'{o}mez}
\affiliation{Center for Applied Structural Discovery, Biodesign Institute, 
Arizona State University, Tempe, Arizona  85287, USA}

\author{Enrique L\'{o}pez-Cabarcos}
\affiliation{Department of Chemistry in Pharmaceutical Sciences, Complutense University of Madrid, E-28040 Madrid, Spain}

\author{Eduardo Cabrera-Granado}
\affiliation{Department of Optics, Complutense University of Madrid, E-28037 Madrid, Spain}

\author{Elena D\'{i}az}
\affiliation{GISC, Department of Materials Physics, Complutense University of Madrid, E-28040 Madrid, Spain}

\author{Jorge Rubio-Retama}
\affiliation{Department of Chemistry in Pharmaceutical Sciences, Complutense University of Madrid, E-28040 Madrid, Spain}
\email{bjrubio@ucm.es}

\title[An \textsf{achemso} demo]
  {FRET distance dependence from upconverting nanoparticles to quantum dots}
 
\abbreviations{FRET, QD, UCNP}

\begin{document}

\begin{tocentry}

\includegraphics[width=9cm]{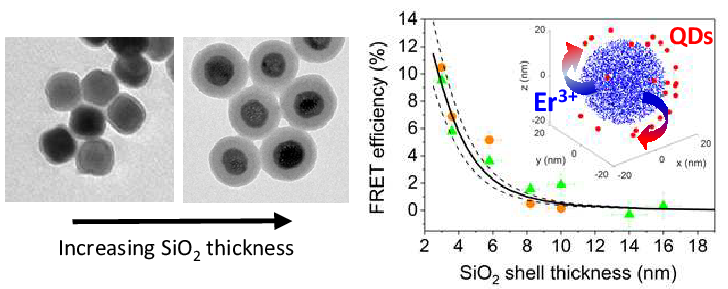}

\end{tocentry}

\begin{abstract}
F\"orster resonant energy transfer (FRET) with upconverting nanoparticles (UCNPs) as donors and quantum dots (QDs) as acceptors has been regarded as a promising tool for biosensing applications. In this work, we use time-resolved fluorescence spectroscopy to analyze the UCNP-to-QD FRET and we focus on the most relevant parameter of the FRET phenomenon,  UCNP-QD distance. This distance is controlled by a nanometric silica shell  around the UCNP surface. We theoretically reproduce the experimental results applying FRET theory to the distribution of emitting erbium ions in the UCNP. This simple model allows us to estimate the contribution of every erbium ion to the final FRET response and to explore different strategies to improve FRET efficiency.
\end{abstract}

\section{Introduction}

In 1948, Theodor F\"orster demonstrated the existence of an electrodynamic phenomenon whereby the energy of a chromophore in the excited state (donor) can be transferred to a neighbor molecule that is in the ground state (acceptor), over a distance larger than the collisional radii. This phenomenon occurs when the donor and the acceptor are in close proximity allowing the energy to be transferred without generation of any photons, through long-range dipole-dipole interactions.\cite{Forster1948} The rate of the energy transfer depends on many different parameters, such as the extent of the spectral overlap between the emission spectrum of the donor and the absorption spectrum of the acceptor, the quantum yield of the donor, the relative orientation of the donor regarding the acceptor transition dipoles, and the distance between the donor and the acceptor.\cite{Lakowicz2006} The FRET's distance dependency has been crucial to detect  two species  in close proximity,  enabling  molecular rulers that can measure the distance between two specific molecules.\cite{Stryer1967} Such a feature in combination with the wide range of techniques for  molecules and  biomolecule functionalization with organic dyes has been applied to study the interaction between DNA, antigens, and proteins in vivo at the nanometer scale.\cite{Byrne2013}
    
During the last decades, advances in the synthesis of upconverting fluorescent nanoparticles with outstanding fluorescence properties and higher photochemical stability than that of classical organic dyes have paved the way for analytical platforms using UCNP-based FRET systems.~\cite{Riuttamaki2011,Vetrone2010,kuningas2007,Jiang2010,Chen2012,Ding2015,Huang2016,DaCosta2014} Lanthanide doped nanoparticles are a special type of materials that are able to absorb low energy photons, typically in the NIR wavelength, and emit UV-Vis photons. Samples can be illuminated with  high power NIR densities, with a high light penetration depth, and without the risk of degradation or photobleaching, which is of especial interest for \textit{in vivo} analysis.~\cite{DaCosta2014,Liu2014,Nyk2008,Zhou2012}  Furthermore, the large anti-Stokes shift and the long lifetime allows to separate the upconverting luminescence signal from biological sample autofluorescence, thus removing the background noise almost completely. The narrow absorption and emission bands of  upconverting nanoparticles simplify the instrumentation for detecting and analyzing FRET processes.~\cite{Diego2017} However, the quantum efficiency in upconversion processes is rather low and therefore it is necessary to use as acceptor a fluorophore with high quantum yield and strong spectral overlapping. Such requirements are fulfilled by quantum dots, which exhibit superb quantum yield and  tunable absorption bands that can be adjusted to overlap with the emission band of the donor. Thus, the use of lanthanide doped nanoparticles as donors and quantum dots as acceptors can be considered as ideal  F\"orster pairs for bioassays.~\cite{Charbonniere2006,Doughan2014,Cui2015,Hildebrandt2017} Recently, Mattsson et al. have demonstrated a mix-and-measure UCNP-to-QD FRET system for rapid homogeneous bioassays for detecting analytes in aqueous solutions at  nanomolar concentrations.~\cite{Mattsson2015} 

A fundamental study of FRET  between NaYF$_4$:Er$^{3+}/$Yb$^{3+}$ nanoparticles and CdSe QDs was reported by Bednarkiewicz et al.~\cite{Bednarkiewicz2010} They observed an Er$^{3+}$ upconversion fluorescence lifetime decrease from 153 $\mu$s to 130 $\mu$s due to the presence of QDs. These authors  suggested two strategies to increase FRET efficiency: 1) diminishing the size of the UCNPs and 2) distributing the lanthanide active ions only in the outer shell of a core/shell UCNP. In this context, Murh et al. have recently analyzed the efficiency of FRET from UCNPs to organic dyes for UCNPs of different sizes.~\cite{Muhr2017} As the UCNP size decreases, two competing phenomena take place. On the one hand, an increasing percentage of the UCNP Er$^{3+}$ ions becomes involved in FRET, increasing its efficiency. But on the other hand, the surface-to-volume ratio increase leads to more luminescence quenching at the UCNP surface, which reduces  FRET strength. They indeed found a maximum FRET efficiency for a 21 nm-sized UCNP. In addition, Bhuckory et al. have analyzed the efficiency of FRET between Er$^{3+}$ ions distributed over either the core or the shell of  UCNP and Cy3.5 dye acceptors.~\cite{Bhuckory2017} The authors use the FRET efficiency results to estimate parameters such as donor-acceptor distances and Er$^{3+}$ donor quantum yields. Very recently, Marin et al. \cite{Marin2018} have experimentally demonstrated  enhanced FRET from LiYF$_4$:Yb$^{3+}$,Tm$^{3+}$ UCNPs to CuInS$_2$ QDs by following two strategies: reducing the size of the UCNPs and doping the lanthanide active ions in the outer shell of a core/shell particle. 

Despite these interesting works in UCNP-based FRET systems, to the best of our knowledge, no exhaustive studies of the FRET efficiency dependence on distance have been carried out. These systems are rather complex, because the potential donors are distributed inside the UCNPs, which sizes are comparable to the length scale of FRET. This leads to a distribution of donor-acceptor distances, a crucial key to the energy transfer phenomena. In this work we present an in-deep study of the distance dependence of FRET in UCNP-QD systems with the aim of developing more efficient UCNP-QD FRET biosensors. To control the UCNP-QD distance, we cover the UCNPs with a nanometric silica shell. By varying the shell thickness, we analyze the change of the upconversion fluorescence lifetime and, therefore, the FRET efficiency. The experimental results are interpreted taking into account the multiple distances between individual donor ions inside the UCNP and the FRET acceptors on its surface.
	
\section{Experimental Section}

\subsection{Chemicals}

ErCl$_{3}$·6H$_{2}$O 99.9\%, YbCl$_{3}$·6H$_{2}$O 99.998\%, YCl$_{3}$·6H$_{2}$O 99.99\%, oleic acid (OA) technical grade 90\%, NH$_{4}$F 98\%, NaOH 98\%, methanol 99.9\%, anhydrous N,N-dimethyl formamide 99.8\%, 1-octadecene (1-ODE) technical grade 90\%, n-hexane 97\%, tetraethyl orthosilicate 98\%, succinic anhydride 99\%, NH$_{4}$OH ACS reagent 28-30\%, IGEPAL CO-520, (3-aminopropyl)-triethoxysilane (APTES) 99\%, and CdTe core-type quantum dots COOH functionalized (part number 777943) were acquired from Sigma-Aldrich and used as received.

\subsection{Synthesis of NaY$_{0.78}$F$_{4}$:Yb$_{0.2}$,Er$_{0.02}$ } 

The synthesis of monodisperse UCNPs with a composition of NaY$_{0.78}$F$_{4}$:Yb$_{0.2}$,Er$_{0.02}$ was performed by using the oleate route, which was first reported by Li and Zhang. \cite{Li2008} Briefly, 15 mL of 1-ODE and 6 mL of OA were mixed in a 100 mL round bottom flask with 3 necks. 233 mg of YCl$_{3}$·6H$_{2}$O (0.78 mmol), 78 mg of YbCl$_{3}$·6H$_{2}$O (0.2 mmol), and 7.9 mg of ErCl$_{3}$·6H$_{2}$O (0.02 mmol) were mixed with the previous solution and heated up at 140 $^{o}$C for 1 hour under a constant magnetic stirring and N$_{2}$ flow to ensure the complete dissolution of the rare earths and obtain a transparent solution with a yellowish color. After this time, the solution is cooled down at room temperature and a fresh methanol solution is prepared by dissolving 100 mg of NaOH (2.5 mmol) and 148 mg of NH$_{4}$F (4 mmol). The methanol solution with NaOH and NH$_{4}$F is added dropwise to the rare earth solution and left stirring for 30 minutes at room temperature. Finally the temperature is raised at 110 $^{o}$C with a constant N$_{2}$ flow, kept for 15 minutes at this temperature and for other 10 minutes under vacuum to ensure the complete evaporation of methanol and water. The temperature of the resulting solution is then raised to 330 $^{o}$C and kept at this temperature for 60 minutes. After this time, the reaction was quickly cooled down to room temperature. The UCNPs were separated by centrifugation: we split the liquid in four different centrifuge tubes and filled them with a solution 1/1 of H$_{2}$O and ethanol. Then, we centrifuged the tubes at 8000 rpm for 10 minutes. The supernatant was discarded and the nanoparticles re-suspended in 0.5 mL of n-hexane and then re-precipitated with ethanol and centrifuged again at 8000 rpm for 10 minutes. Finally the UCNPs were re-suspended in 10 mL of n-hexane with a concentration of 12 g/L and stored for further experiments.

\subsection{Preparation of NaY$_{0.78}$F$_{4}$:Yb$_{0.2}$,Er$_{0.02}$@SiO$_{2}$-NH$_{2}$} 

The UCNPs were covered with different thickness of SiO$_2$ following the reverse microemulsion method. \cite{Marco2016,Zhelev2006} 240 $\mu$L of IGEPAL CO-520 were mixed with 4 mL of n-hexane, 1 mL of UCNPs dispersed in the previous solution (12 g/L) and 40 uL of NH$_{4}$OH. The resulting solution was submitted to sonication until a transparent emulsion was formed. Different volumes of TEOS ranging from 10 to 30 $\mu$L were added depending on the desired SiO$_{2}$ shell thickness. The reaction was stopped after 18 hours with the addition of methanol to disrupt the microemulsion, then it was centrifuged at 9000 rpm for 10 minutes, and the supernatant was discarded. This process was repeated 3 times to remove the excess of IGEPAL CO-520. Finally the SiO$_{2}$ surface was functionalized with the amino groups by the addition of APTES (15 $\mu$L, 0.068 mmol) to the synthesized UCNPs@SiO$_{2}$ dispersed in 5 mL of ethanol. The heterogeneous solution was stirred overnight at room temperature, and the next day the UCNPs@SiO$_{2}$-NH$_{2}$ were recovered by centrifugation at 9000 rpm for 10 minutes. The supernatant was discarded and the nanoparticles were re-dispersed in ethanol. This process was repeated three times.

\subsection{Morphological characterization}

The electron microscopy images were acquired in transmission mode (TEM) using a JEOL JEM 1010 microscope operated at 80 kV (JEOL, Japan; 80 kV) equipped with a digital camera Gatan megaview II.
High Resolution TEM (HRTEM) and Low Angle Annular Dark Field Scanning TEM (LAADF-STEM) images were acquired using a JEOL JEM 3000F operated at 300 kV and a Gatan ADF detector.

\subsection{Optical characterization}

Absorbance spectra of QDs water solutions at different concentrations were measured using a UV-VIS spectrometer (\textit{Ocean Optics}, RedTide 650) with 3 mm path length cuvettes. 

Upconversion fluorescence spectra were recorded with a fluorescence home-built system. The excitation beam comes from a pigtailed 10W CW laser (JDSU, L4-9897603) working at 976 nm and provided with a current and temperature controller (\textit{ILX Lightwave}, LDX-36025-12 and LDT-5525B, respectively). The laser beam is transmitted through a long-pass dichroic filter (\textit{Semrock}, FF757-Di01) and then focused on the sample with a 10X objective. The upconversion photoluminescence coming from the sample is reflected by the dichroic mirror towards a short-pass filter which blocks the IR reflected radiation (\textit{Semrock}, FF01-775/SP). Then, it is focused into an optical fiber connected to a monochromator (\textit{Horiba Jobin Yvon}, iHR320). The monochromator is equipped with a photomultiplier tube (\textit{Hamamatsu}, R928) and uses a 1200 gr/mm grating blazed at 900 nm. 
In order to characterize the laser intensity in the sample, we measure the laser power with a thermal sensor power meter (\textit{Thorlabs}, S310C) and the beam size (HWHM) using the slit scan technique \cite{McCally1984} being this size around 150~$\mu$m. In our measurements, we use excitation laser powers around 1.5 W, leading to laser intensities around 2 kW/cm$^2$. This allows us to ensure that the laser operates below the excitation saturation intensity of the transition $^{2}$F$_{7/2}$ $\rightarrow$ $^{2}$F$_{5/2}$ for the Yb$^{3+}$ ions (see Figure \ref{fig:AbsorbanceQD_UCfluorescence}B), which is $I_{sat} = \hbar \omega / (2 \sigma \tau_{Yb}) = 3$ kW/cm$^2$, where $\tau_{Yb}$=2 ms is the excited level lifetime, and $\sigma=1.7 \times 10^{-20}$ cm$^2$ is the absorption cross-section.

\subsection{Upconversion emission lifetime measurements}

Fluorescence lifetimes were measured using the time-resolved photon counting method. Laser current is pulsed to generate excitation pulses of 40 $\mu$s with repetition rate of 125 Hz. The fluorescence emission at 540 nm is detected by the PMT which is directly connected (without using a pre-amplifier) to a 50 $\Omega$ input of a digital oscilloscope (\textit{Agilent}, DSO9104A). The signal from the laser current controller is used to trigger the oscilloscope. We developed a Matlab program that analyzes directly in the oscilloscope each recorded signal in real-time. This code simulates the discriminator and the multichannel counter \cite{Stacewicz1997}. Upon analysis of more than 5000 trigger signals, we obtain a fluorescence decay curve. Decay curve measurements were repeated at least three times for each sample under the same experimental conditions.

\section{Results and discussion}

\begin{figure}[ht]
	\centering
		\includegraphics[width=13.5cm]{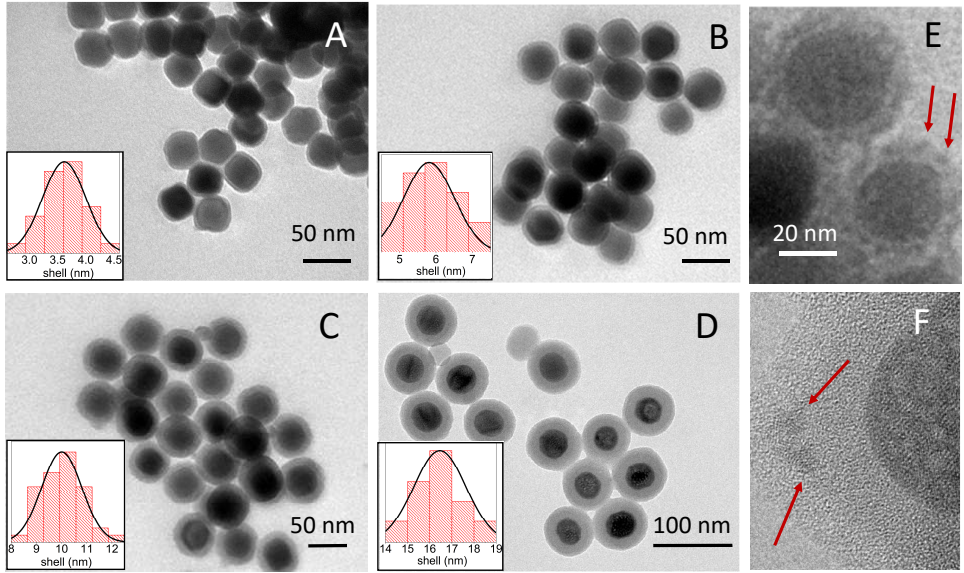}
		\caption{TEM micrographs of the synthesized NaYF$_4$;Yb,Er@SiO$_2$ nanoparticles with different SiO$_2$ shell thickness: A) (3.6 $\pm$ 0.4) nm, B) (5.8 $\pm$ 0.7) nm, C) (10.0 $\pm$ 0.8) nm, and D) (16 $\pm$ 1) nm. E) TEM  and F) HRTEM  micrographs after attaching CdTe QDs (marked with arrows) on the surface of the NaYF$_4$;Yb,Er@SiO$_2$ nanoparticles.  } \label{fig:TEM}
\end{figure}

To study the distance dependence of energy transfer between UCNP and QD pairs, we covered the surface of the UCNPs with a controlled thickness, uniform, amorphous silica layer. TEM micrographs from the as-synthesized UCNPs@SiO$_{2}$-NH$_2$ show highly monodisperse UCNPs with a mean diameter of (33 $\pm$ 3) nm (see Figure \ref{fig:TEM}). We varied the SiO$_2$ shell thickness from 3 nm to 16 nm. As an example, Figure \ref{fig:TEM} shows TEM images for UCNPs@SiO$_{2}$-NH$_2$ with different silica shell thickness: A)  3.6 nm, B) 5.8 nm, C) 10 nm, and D) 16 nm. After surface modification with amine groups, the NaYF$_4$:Yb,Er@SiO$_2$-NH$_2$ nanoparticles have a z-potential of +22 mV. CdTe QDs have an average diameter of 3 nm. QDs are functionalized with carboxylic groups and have a z-potential of -27 mV, so strong electrostatic interaction between positively charged UCNPs and negatively charged QDs will occur.


We prepared a solution containing UCNPs@SiO$_{2}$-NH$_2$ and CdTe-COOH QDs. First, the UCNPs@SiO$_{2}$-NH$_2$ were dispersed in ethanol at 5 g/L, while the CdTe-COOH QDs were dispersed in distilled water at 2.5 g/L. Then, we mixed equal volumes of the ethanol solution with the UCNPs@SiO$_{2}$-NH$_2$ and the aqueous solution with the CdTe-COOH QDs, so that an excess of QDs was used. Negatively charged QDs were electrostatically absorbed on the surface of positively charged UCNPs, as shown in the TEM and HRTEM images of Figures \ref{fig:TEM}E and \ref{fig:TEM}F, respectively. Single drops of this mixture were placed on a filter paper and allowed to dry at room temperature in air. The filter paper containing the mixture was set between two microscope glass slides for optical characterization.

As a reference, in the same way we prepared a filter paper with a drop of the ethanol solution containing UCNPs@SiO$_{2}$-NH$_2$ and  distilled water without QDs  mixed in equal volumes (hereafter named series I). In this reference sample we use the same UCNP concentration than that of the sample with both UCNPs and QDs, while avoiding possible changes in the contributions to fluorescence quenching due to water \cite{Riuttamaki2011}. In a second experiment (hereafter referred to as series II), the reference sample is made by using only the ethanol solution containing UCNPs@SiO$_{2}$-NH$_2$. In this last case, the possible effect of water was accounted for  by taking into consideration that energy transfer between UCNPs and QDs should be negligible at very large UCNP-QD distances. This allows us to test the robustness of our experiments. 

\begin{figure}[ht]
	\centering
	\includegraphics[width=12cm]{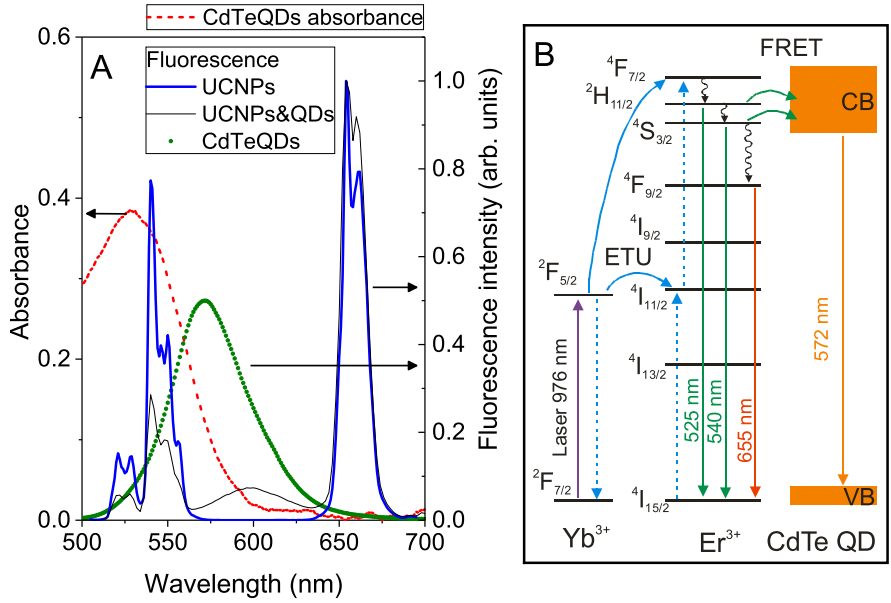}
	\caption{A) (Left axis) Absorbance spectrum for 0.5 g/L water solution of CdTe QDs in a 3 mm cuvette (red curve). (Right axis) Fluorescence spectrum for: CdTe QDs water solution when excited at 455nm (green curve), UCNPs@SiO$_{2}$ ethanol solution when excited at 976 nm (blue curve), and UCNPs@SiO$_{2}$ (3 nm-shell) with QDs dry sample when excited at 976 nm (black curve). B) Energy transfer upconversion (ETU) scheme for populating the green emission levels 
$^{2}$H$_{11/2}$ and  $^{4}$S$_{3/2}$ by a 976 nm laser and FRET scheme from these levels to the CdTe QD.} \label{fig:AbsorbanceQD_UCfluorescence}
\end{figure}

Figure~\ref{fig:AbsorbanceQD_UCfluorescence}A shows the optical characterization for the as-synthesized donors (UCNPs) and the acceptors (QDs) used in the FRET system under consideration. 

The upconversion fluorescence emission spectra of NaYF$_4$:Yb,Er@SiO$_2$ nanoparticles under a 976 nm CW excitation laser is shown in 
Figure~\ref{fig:AbsorbanceQD_UCfluorescence}A (blue curve, right axis). Two green emission peaks near 525 nm and 540 nm are observed. These peaks correspond to $^{2}$H$_{11/2}$ $\rightarrow$ $^{4}$I$_{15/2}$ and $^{4}$S$_{3/2}$ $\rightarrow$ $^{4}$I$_{15/2}$ transitions of the Er$^{3+}$ ions, respectively (see in Figure \ref{fig:AbsorbanceQD_UCfluorescence}B the energy transfer upconversion mechanism populating these two levels). A red emission, with similar intensity, near 655 nm is also observed due to $^{4}$F$_{9/2}$ $\rightarrow$ $^{4}$I$_{15/2}$ transition (see Figure \ref{fig:AbsorbanceQD_UCfluorescence}B).  Figure~\ref{fig:AbsorbanceQD_UCfluorescence}A also shows the absorbance of a CdTe QDs water solution with 0.5 g/L in a 3 mm path length cuvette (red curve, left axis). 
Notice that the size of the QDs can be inferred from the position of their absorption peak \cite{Yu2003,Dagtepe2007}. In our case, we found an absorption peak around 529 nm, that gives us an estimation of the QDs size of 3 nm, in good agreement with that obtained from High-Resolution TEM images.
Therefore, since the green emission bands of the UCNPs (donors) perfectly overlap with the absorption peak of the QDs (acceptors), the possibility of an efficient nonradiative energy transfer from UCNPs to QDs is ensured provided that both are close enough. Additionally, the QDs fluorescence emission spectra under excitation by a LED radiation at 455 nm is shown in Figure~\ref{fig:AbsorbanceQD_UCfluorescence}A (green curve, right axis). The QDs orange emission band whose peak is around 572 nm  does not overlap with the emission bands of the Er$^{3+}$ ions avoiding mixed fluorescent emissions. 

Now, let us test the interaction between UCNPs and QDs. Figure~\ref{fig:AbsorbanceQD_UCfluorescence}A shows the fluorescence emission spectrum from the dry sample containing UCNPs with QDs electrostatically absorbed on their surface (black curve, right axis). There, a new emission peak around 600 nm and a reduction of the green emission bands of the UCNPs are observed. Both features reveal that QDs are being excited by the $^{4}$S$_{3/2}$, $^{2}$H$_{11/2}$ $\rightarrow$ $^{4}$I$_{15/2}$ transitions of the Er$^{3+}$ ions. Note that QDs cannot be excited with the IR radiation that excites the UCNPs. The emission peak arising from the QDs fluorescence appears as red-shifted respect to that obtained for the isolated QDs solution excited under UV light. A similar shift was previously reported as related to the inner filter effect due to the increase of particle concentration during the drying process \cite{Bednarkiewicz2010}.

Since the reduction of the UCNPs green fluorescence intensity is due to both non-resonant energy transfer and re-absorption,  to characterize in detail the  Er$^{3+}$ to QD FRET mechanism, luminescence lifetime studies are needed. Thus, fluorescence decay signals for Er$^{3+}$ ions at 540 nm ($^{4}$S$_{3/2}$ $\rightarrow$ $^{4}$I$_{15/2}$) were measured. 
\begin{figure}[ht]
	\centering
	\includegraphics[width=8.5cm]{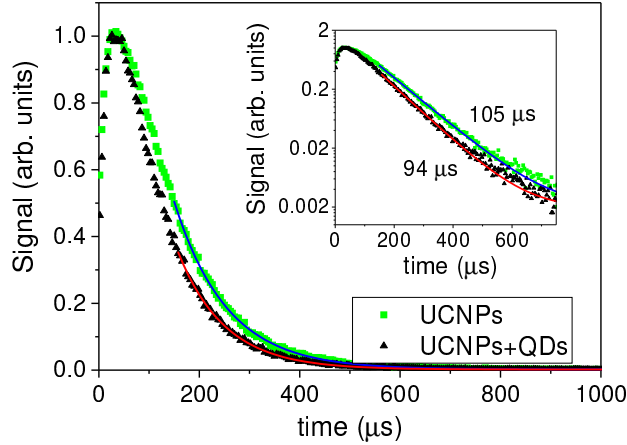}
	\caption{Normalized fluorescence decay curves at 540 nm (transition $^{4}$S$_{3/2}$ $\rightarrow$ $^{4}$I$_{15/2}$) in presence (black) or absence (green, series I) of CdTe QDs for UCNPs with a SiO$_2$ shell of 3 nm. Excitation laser at 976 nm with 40 $\mu$s pulses at 125 Hz. Inset: zoom of previous curves in a semilog scale.}
\label{fig:decaytime}
\end{figure}
As an example, Figure \ref{fig:decaytime} shows the fluorescence intensity decay of UCNPs with a 3 nm silica shell, in presence (black) and absence (green) of QDs. Fluorescence signals show an initial increase that is a signature of upconversion process energy transfer from Yb$^{3+}$ to Er$^{3+}$ ions within the UCNPs, followed by an exponential decay. The fluorescence decay time was obtained by fitting the decay curve to a single exponential function. For the fitting, we considered  initial time values corresponding to 70\% to 30\% of the maximum decay curve amplitude, while the final fitting time was set to 1.2 ms. This fitting procedure gives us an average lifetime with its standard error. As shown in Figure \ref{fig:decaytime}, the presence of the QDs reduces the  fluorescence lifetime from 104.7 $\pm$ 0.3 $\mu$s to 93.7 $\pm$ 0.2 $\mu$s. This decrease confirms the occurrence of non-radiative energy transfer from UCNPs to QDs. We have also corroborated that this energy transfer process produces a slow component ($<$100 $\mu$s) in the decay of the QD fluorescence at 600 nm, which otherwise would typically be in the nanoseconds range.

Once the FRET mechanism has been verified, its characterization is in order. FRET efficiency ($E$) can be computed from the experimental fluorescence decay curves obtained in Figure \ref{fig:decaytime} as:
\begin{equation}  
\label{eq:FRETexp}
E  = 1 - \frac{\tau_{DA}}{\tau_D},
\end{equation}
where  $\tau_D$  and $\tau_{DA}$ are the donor excited-state lifetime in the absence and in the presence of acceptor, respectively. For the particular case shown in Figure \ref{fig:decaytime}, an efficiency of $E$ = (10.5 $\pm$ 0.4) $\%$ was achieved with a SiO$_2$ shell of 3 nm, which is quite high considering the  relatively large diameter of the UCNPs (33 nm). Note that while fluorescence emission comes from the Er$^{3+}$ ions distributed within the entire nanoparticle, ions far from the particle surface cannot participate in FRET.\cite{Mattsson2015}

\begin{figure}[ht]
	\centering
\includegraphics[width=10cm]{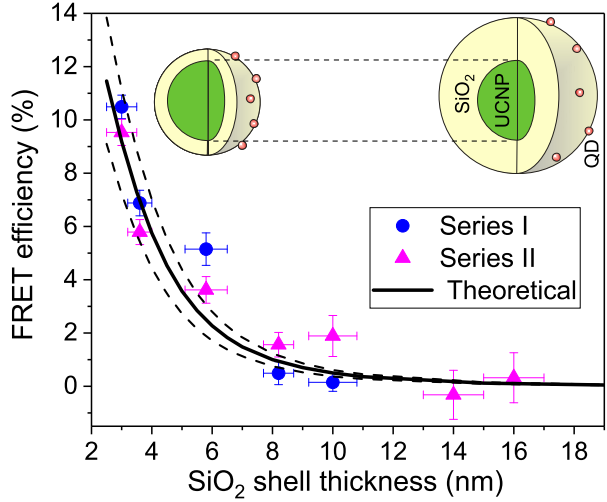}
	\caption{FRET efficiency, $E$, for different silica shell thicknesses on UCNPs with a diameter of 33 nm. Laser intensity: 1.7 kW/cm$^2$ (triangles), 2.4 kW/cm$^2$ (circles). Theoretical prediction of FRET efficiency (Eq. \ref{eq:AveEff}) for a F\"orster distance of $R_0 = 5.5$ nm and a number of QDs of 24 (solid line). Dashed lines correspond to a variation on the number of QDs of $\pm $ 6.  
	} \label{fig:FRET}
\end{figure}

To analyze the behavior of the F\"orster resonance energy transfer with the distance between the UCNPs and the QDs, we measured the fluorescence decay times of UCNPs with different SiO$_2$ shell thicknesses, with and without QDs. The FRET efficiency as a function of shell thickness is plotted in Figure \ref{fig:FRET}.  The general trend is a fast decrease of the FRET efficiency for donor-acceptor distances up to 12 nm, above which the energy transfer is negligible. A maximum FRET efficiency of around 10\% is obtained, in agreement with the values reported in previous works\cite{Bednarkiewicz2010}. We obtained similar results for the range of excitation laser intensities used in the experiments and also for the two reference samples, series I and II. 

Notice that our measurements do not show saturation of  $E$ as the silica shell thickness decreases, as it would be the case when dealing with a single donor-acceptor pair. This feature is directly related to the fact of having a distribution of multiple donors (Er$^{3+}$ ions) inside the UCNP as we will see later. We can estimate the F\"orster radius $R_0$ as the distance where half of the maximum efficiency is measured.  As shown in Figure \ref{fig:FRET},  $R_0$ $\simeq$ 6 nm.

\subsection{Theoretical FRET efficiency}
 
After analyzing the experimental data, let us now have a look at the theoretical counterpart for further interpretation of our results. 
The rate of energy transfer from a single donor to a single acceptor separated by a distance $R$ is given by \cite{Lakowicz2006}

\begin{equation}
   k_{ET} = \frac{9 (ln10)}{128 \pi^5 N_A n^4} \; \frac{\eta_D \kappa^2}{\tau_D R^6} \;  J
\ ,
\qquad
J =  \int_{0}^{\infty} F_D(\lambda) \epsilon_A(\lambda) \lambda^4 d\lambda \ ,
\label{eq:kT}
\end{equation}

\noindent  
where $N_A$ is the Avogadro's number, $n$ is the refraction index of the medium surrounding the FRET pair, $\tau_D$ is the donor excited-state lifetime, and $\eta_D$ is the intrinsic quantum yield of the donor in the absence of acceptor; that is, the quantum yield of the Er$^{3+}$ ion excited-state \cite{Bhuckory2017}. Note that $\eta_D = \tau_D / \tau_D^{rad}$, being $\tau_D^{rad}$ the radiative donor lifetime.
Furthermore, $\kappa^2$ is a factor describing the relative orientation of the donor and the acceptor dipole moments, and the integral $J$ measures the overlap between the donor emission spectrum and the acceptor absorption spectrum. Last, $F_D(\lambda)$ is the fluorescence intensity of the donor normalized to the total intensity (i.e., $\int F_D(\lambda) d\lambda = 1$), and $\epsilon_A(\lambda)$ is the molar extinction coefficient of the acceptor.
In addition, the rate of energy transfer can be rewritten as 
\begin{equation}  \label{eq:kT2}
k_{ET} = \frac{1}{\tau_D} \Big(\frac{R_0}{R}\Big)^6 ,
\end{equation}
where $R_0$ is the F\"orster distance  defined as the distance at which half of the donor ions decay by transferring their energy to the acceptors; that is, the distance at which the energy transfer rate is equal to the donor decay rate in the absence of acceptor.
By comparing Eqs. \ref{eq:kT} and \ref{eq:kT2}, one can obtain a closed expression for $R_0$ as follows:
\begin{equation} \label{eq:R0_Lakowicz}
R_0 = 0.0211 \left( \kappa^2 \eta_D  n^{-4} J\right)^{1/6}  \; 
[{\rm nm}] \; ,
\end{equation}

\noindent where J is evaluated with the wavelength expressed in [nm] and the molar extinction coefficient in [M$^{-1}$ cm$^{-1}$]. 

\begin{figure}[ht]
	\centering
    \includegraphics[width=8.5cm]{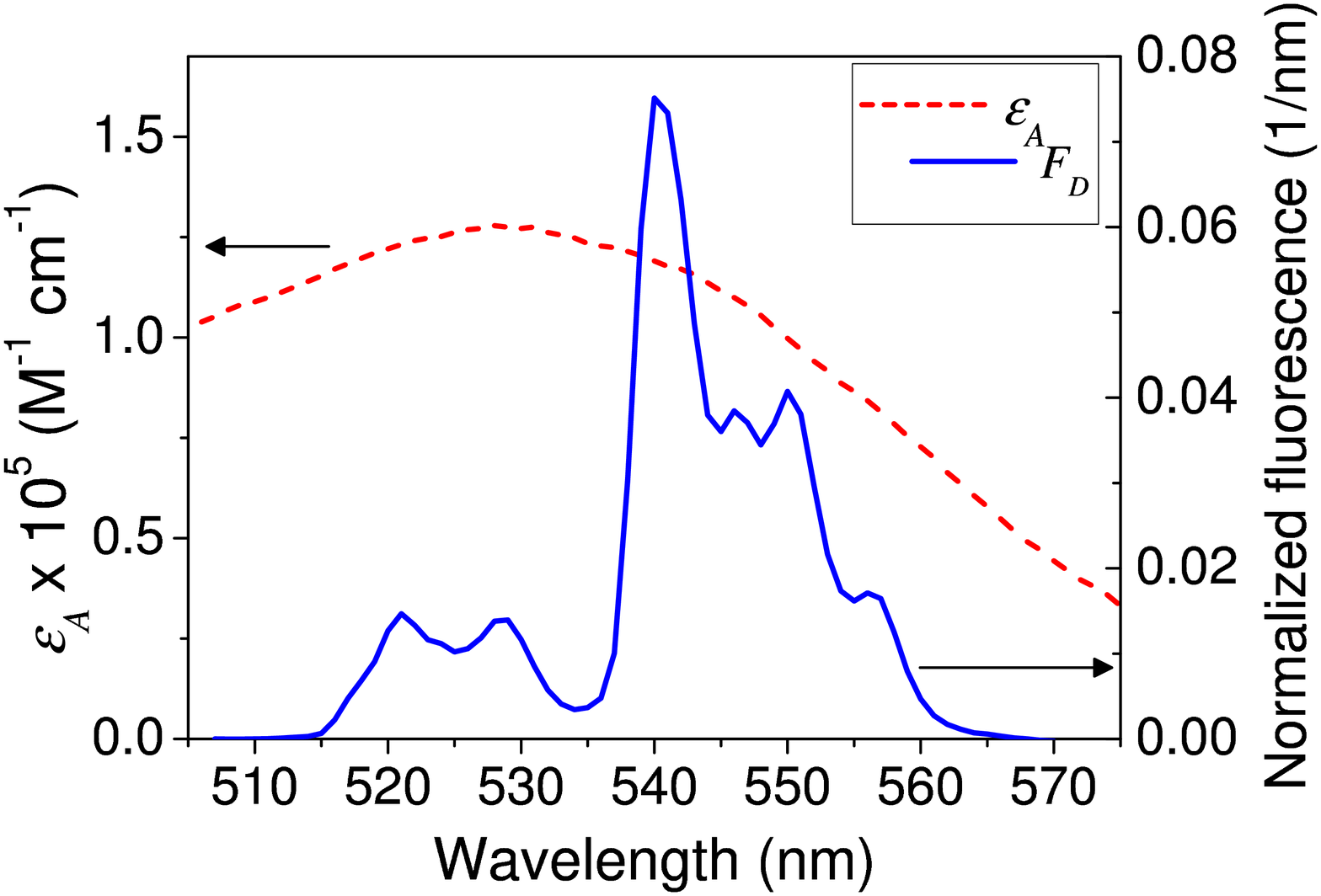}
	\caption{
		(Left axis) Molar extinction coefficient, $\epsilon_A(\lambda)$, of a $10^{-5}$ M QDs water solution in a cuvette with 3 mm path length. (Right axis) Upconversion fluorescence intensity spectrum, $F_D(\lambda)$, (normalized to unity area under the curve) used to calculate the overlap integral $J$. Both represented as a function of the excitation wavelength $\lambda$.} \label{fig:sizeQDandJ}
\end{figure}

In order to estimate the F\"orster distance $R_0$, 
we calculate the overlap integral $J$ from the experimental measurements of $\epsilon_A(\lambda)$ and $F_D(\lambda)$. The molar extinction coefficient of the QDs is obtained through the Beer-Lambert's law by measuring the absorbance at different concentrations and taking into account the cuvette path length ($L=3$ mm). The result is shown in Figure~\ref{fig:sizeQDandJ}. From the experimental data shown in Figure \ref{fig:sizeQDandJ}, the overlap integral results to be $J = 9.6 \times 10^{15}$ M$^{-1}$ cm$^{-1}$ nm$^4$. Note that this value can be roughly estimated by considering a nearly constant value of $\epsilon_A$ in the green-region of the spectrum around $\lambda \simeq 540$ nm, which gives  $J \simeq \epsilon_A \lambda^4 \simeq  10^{16}$ M$^{-1}$ cm$^{-1}$ nm$^4$.

Other important parameters necessary to evaluate Eq. \ref{eq:R0_Lakowicz} are $\kappa$, $n$ and $\eta_D$. The orientation factor of donor-acceptor dipoles $\kappa^2$, is usually assumed to be 2/3; which corresponds to the dynamic random averaging. The material through which FRET mainly takes place is silica whose index of refraction is $n=1.46$. Last, the intrinsic quantum yield $\eta_D$, which measures the probability that donor de-excites radiatively in the absence of acceptor, has been estimated  as around 20-30\% by Bhuckory et al. from FRET measurements of  UCNPs with different core-shell architectures.\cite{Bhuckory2017} Note that this value is much larger than the one corresponding to the overall UCNP quantum yield. We estimate the intrinsic quantum yield by considering that the radiative lifetime for the excited-donor level ($^4S_{3/2}$) is in the millisecond range $\tau_D^{rad}=0.5$ ms,
\cite{Weber1968}
and by using our experimental fluorescence decay rate  $\tau_D \simeq 105$ $\mu$s, which leads to  $\eta_D = \tau_D / \tau_D^{rad}\simeq 0.21$. For the sake of simplicity, we assume the quantum yield to be the same for all the UCNPs@SiO$_2$. This is reasonable since the donor lifetime obtained without QDs was quite similar throughout the range of silica shell thicknesses used in our experiments. 
Finally, by evaluating Eq. \ref{eq:R0_Lakowicz} with these parameters, our theoretical prediction for the F\"orster distance is $R_0 \simeq 5.5$ nm, which is in very good agreement with the experimental value shown on Figure   \ref{fig:FRET}.  

With this estimation in mind, to understand the observed dependency of the  FRET efficiency on distance, we must consider the distribution of multiple donor-acceptor pairs. First, let us express the transfer efficiency (that is, the probability of de-excitation via energy transfer) of a single donor-acceptor pair (DA) as:
\begin{equation} 
E(k_{ET}^{DA})  = \frac{k_{ET}^{DA}}{ 1 / \tau_D + k_{ET}^{DA}}\ .
\label{eq:DAEff}
\end{equation}

We consider that the energy transfer rate from each Er$^{3+}$ ion inside the UCNP to all the QDs absorbed onto its surface results from the sum of the single  pair transfer rates: $k_{ET}^{D}=\sum_{A=1}^{N_{QD}}k_{ET}^{DA}(A)$, where $N_{QD}$ is the total number of QD acceptors absorbed onto the UCNP surface. 
Here $k_{ET}^{DA}(A)$ results from the evaluation of Eq. \ref{eq:kT2} at the particular donor-acceptor distance R for every case. 
Although we take $N_{QD}$ as a fitting parameter in our simulations, we can roughly estimate this value from the TEM images as $N_{QD}\sim$ 20.
FRET efficiency is then calculated by averaging the efficiency of each Er$^{+3}$ ion inside the nanoparticle, 
\begin{equation}
E=\langle E(k_{ET}^{D})\rangle_{N_{Er}}\ . 
\label{eq:AveEff}
\end{equation}
Here, the number of Er$^{+3}$ ions inside the nanoparticle is $N_{Er} = f_{Er} m_{NP} N_A / W =  4405$, where $f_{Er}=0.019$ is the fraction of Er$^{+3}$ ions, $W=205.3$ g/mol is the molar weight of NaYF$_4$:Yb/Er, and $m_{NP}= 7.9 \times 10^{-17}$ g is the mass of the UCNP.

\begin{figure}[ht]
	\centering
	\includegraphics[width=8.5cm,angle=90]{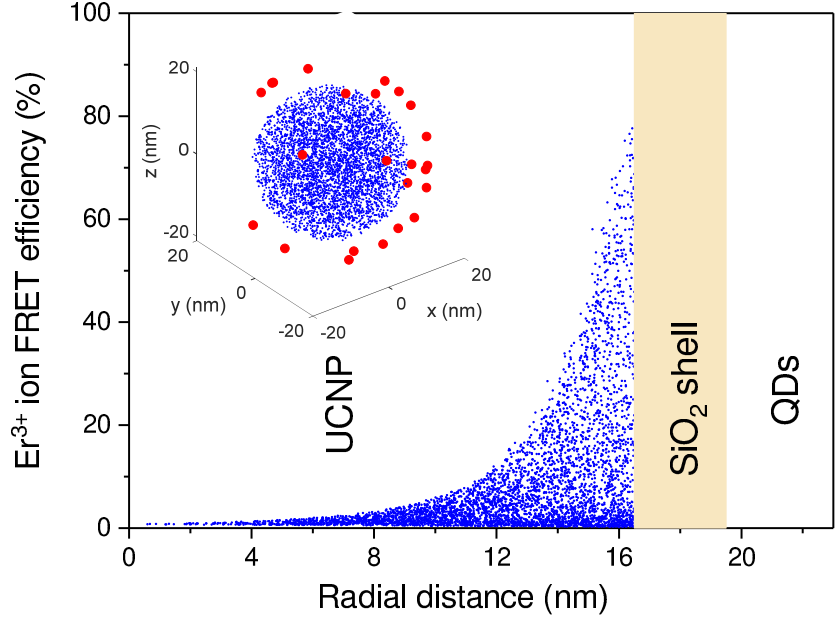}
	\caption{Distribution of the FRET efficiency achieved by each Er$^{3+}$ ion inside an UCNP with 3 nm SiO$_2$ shell, $E(k_{ET}^D)$, as a function of the radial distance from the particle center. Inset: Distribution of: Er$^{3+}$ ions (blue points) inside a 33 nm UCNP with a SiO$_2$ shell of 3 nm, and 24 QDs randomly absorbed onto the particle-shell surface (red points).
	} \label{fig:simulation}
\end{figure}

In our calculations, we generate a uniform random distribution of the  Er$^{+3}$ ions inside the UCNP \cite{Knuth1999}, and we compute the average FRET efficiency using Eq. \ref{eq:AveEff} with the F\"orster distance calculated above, that is $R_0 = 5.5$ nm, and a number of QDs around the particle of 24. Our simulations result presented in Figure \ref{fig:FRET} (solid line) shows a very good agreement with the experimental data. 
We have also included in Figure \ref{fig:FRET} (dashed lines) theoretical curves with a different number of QDs, to account for the observed variation of the number of acceptors per UCNP in TEM images.   

We conclude that the averaging process carried out when considering multiple donor-acceptor configuration is needed in order to explain the FRET efficiency found in the experiments. 
For further analysis of this phenomenon, Figure \ref{fig:simulation} shows the distribution of FRET efficiency for each Er$^{3+}$ ion, $E(k_{ET}^D)$, as a function of its radial distance to the particle center (blue points). Significant energy transfer, larger than 10\%, occurs only for those ions placed farther than 11 nm from the UCNP center. That means that only the ions in close proximity to the UCNP external surface ($\leq$ 5.5 nm) are susceptible to efficiently transfer their energy to the QDs. However, we observe a wide distribution on FRET efficiencies even for these superficial ions since the QDs are randomly distributed on the silica shell without completely covering the UCNP surface (see inset in Figure \ref{fig:simulation}). 

\subsubsection{Optimal design of UCNPs for FRET performance}

As we mentioned in the Introduction, two different strategies have been used to increase FRET efficiency in UCNP systems: reducing the UCNP size and developing inert-core/active-shell UCNP architectures.\cite{Bhuckory2017,Marin2018}  In both cases, the fraction of active ions that can exhibit an efficient energy transfer increases. Marin et al.\cite{Marin2018} have experimentally reported a two-fold FRET efficiency increase by applying these strategies to square-based bipyramidal shape UCNPs and QDs. 

\begin{figure}[ht]
	\centering
    \includegraphics[width=12.5cm]{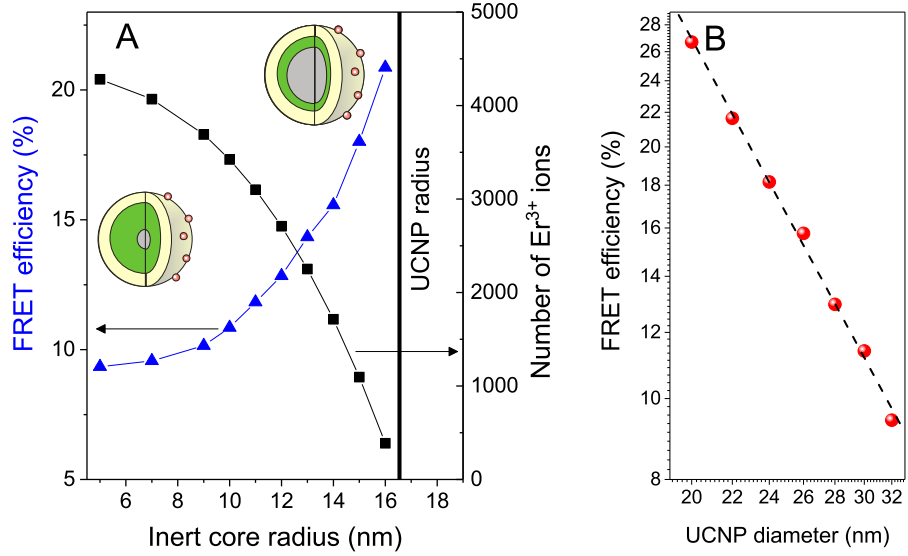}
	\caption{A) FRET efficiency, $E$, (left axis) and total number of Er$^{3+}$ ions inside each UCNP, $N_{Er}$, (right axis) as a function of the inert core radius for a particle with a diameter of 33 nm. B) FRET efficiency, $E$, as a function of the UCNP diameter $d_{NP}$ (symbols). The dashed line is the linear fit in the log-log plot.   
	} \label{fig:coreinactivo}
\end{figure}

Let us use our theoretical approach to estimate the FRET efficiency that could be achieved under these two enhancement strategies for our particular UCNP-QD system. We consider UCNPs as the ones used in the experiments but with a tunable size inert-core, that is, particles with 33 nm of diameter, a silica shell of 3 nm, and we consider 24 QDs absorbed onto the particle surface (see particle scheme in Figure \ref{fig:coreinactivo}A). 
The quantum yield of particles with different inert-core diameters is expected to change. However, determining this magnitude is a challenging experimental problem, and as a first  approximation, we consider here that all particles have roughly the same quantum yield, that is, the same F\"orster distance $R_0 = 5.5$ nm.
Figure \ref{fig:coreinactivo}A shows the FRET efficiency and the number of Er$^{3+}$ ions per nanoparticle as a function of the inert-core radius. We observe that in order to obtain a notable increase of FRET efficiency, an inert-core radius larger than half of the particle radius is required (see blue symbols in Figure \ref{fig:coreinactivo}A, left axis). Note that the reduction in the total number of Er$^{3+}$ ions in a particle with an inert core  is not significant until the inert core takes half of the particle radius (see black symbols in Figure \ref{fig:coreinactivo}A, right axis). We obtain a two-fold FRET efficiency increase  when the inert radius is almost as large as the particle radius, i.e., when the active shell is around 2 nm thick.  

Last, we now analyze the effect of UCNP size on the FRET response. We consider UCNPs with diameter from 32 to 20 nm, all particles with a 3 nm silica shell. The number of QDs absorbed onto the UCNP surface and the donor quantum yield are properties that depend on UCNP size, but, for the sake of simplicity, we consider the values previously used, i.e., $N_{QD}=24$ and $R_0=5.5$ nm. 
Figure \ref{fig:coreinactivo}B shows the FRET efficiency as a function of the UCNP size in a log-log plot. The linear fit shown in this figure (dashed line) indicates that the FRET efficiency follows a power law behavior with the UCNP diameter $d_{NP}$ with an exponent close to  minus 2, that is, $E \propto  d_{NP}^{-2}$. Under these assumptions, the change of FRET efficiency is proportional to the nanoparticle area. For example, by reducing the UCNP diameter from  33 nm  to  23 nm, therefore reducing the nanoparticle volume by 1/3,  we calculate a FRET efficiency of around 20\%, which is similar to our previously calculated efficiency for a 33 nm particle consisting of a 2 nm active-shell and a 31 nm inert-core, and  doubles the value measured in our experiments for the 33 nm diameter UCNPs. Similar increases in efficiency with reductions in particle diameter have been  experimentally reported in a previous work\cite{Marin2018}.

\section{Conclusion}

We have studied the FRET phenomenon from UCNPs to QDs, which is a system with several advantages. The CdTe QDs exhibit a strong absorption band that perfectly overlaps with the fluorescence emission band of the UCNPs. Furthermore, cross-excitation is avoided since the infrared UCNP donor excitation does not produce fluorescence in the QD acceptors.  We experimentally verified FRET between 33-nm diameter, positively-charged NaYF$_4$:Yb,Er@SiO$_2$-NH$_2$ particle donors and 3-nm diameter negatively-charged CdTe-COOH QDs acceptors. In our system, several small CdTe QDs  were absorbed on each larger UCNP@SiO$_2$ surface by electrostatic interaction. This multiple acceptor configuration enhances the FRET response. The QD adsorption leads to a significant reduction of the upconversion fluorescence lifetime, which clearly demonstrates that an additional decay pathway for the excited UCNPs takes place due to non-radiative energy transfer to the QDs. FRET is limited by the fact that the donor and acceptor must be in extreme close proximity. In our system, a  F\"orster distance of $R_0 \simeq 5.5$ nm was observed, which is a large value compared to the F\"orster radius characteristic of organic fluorophores of a few Angstroms to two nanometers, but still small compared to the UCNP radius. We experimentally analyzed the effect of the UCNP-QD distance in the FRET efficiency by varying the thickness of the UCNPs silica shell from 3 to 16 nm. A maximum FRET efficiency of around 10\% is achieved for the 3 nm silica shell, which is a remarkably large value if we take into account the distribution of all the potential donors, i.e., Er$^{3+}$ ions, inside the relatively large UCNP in comparison with the F\"orster distance. The UCNP-to-QD nonradiative interaction vanished for distances above 12 nm.   

We have theoretically explained the experimental results by calculating the FRET efficiency of each single Er$^{3+}$ ion-QD pair and averaging the FRET response of every Er$^{3+}$ ion inside the UCNP. Therefore, in order to theoretically obtain accurate values of the FRET efficiency in UCNP-based FRET systems this averaging process is required. Indeed, we were able to establish that only Er$^{3+}$ ions placed in close proximity to the UCNP surface, in a shell of around 5 nm (similar to the  F\"orster distance), are able to participate in an energy transfer process with $E>$10\%.  

In summary, our work supports that the main physical characteristics governing the UCNP-to-QD energy transfer process is the relative spatial configuration of donors and acceptors. Our results show that the as-proposed UCNP-QD FRET system is a good potential platform for biosensing short biomolecules whose length of interaction is below the estimated F\"orster distance ($\leq$ 6 nm). Nanoparticle systems with alternative architectures, as the ones discussed in this work, can be used to improve the distribution of donor-acceptor distances for more efficient FRET-based biosensing applications.

\begin{acknowledgement}

This work was supported by Universidad Complutense de Madrid and Santander Bank (PR26/16-12B), Ministerio de Econom\'{i}a y Competitividad-MINECO (MAT2016-75955, MAT2017-83111R), and Comunidad de Madrid (B2017/BMD-3867 RENIM-CM). 

\end{acknowledgement}

\bibliography{FRET_Melle}
\end{document}